# Electrophysiological mechanisms in neurodegenerative disorders and role of non-pharmacological interventions in improving neurodegeneration and its clinical correlates: A review

Sheng Mai1
1. University College London, Institute of Cognitive Neuroscience, London, United Kingdom

## ABSTRACT

Mild cognitive impairment (MCI) leading to dementia results in a constellation of psychiatric disorders including depression, mood disorders, schizophrenia and others. With increasing age, mild cognitive impairment leads to increased disability adjusted life-years and healthcare burden. A huge number of drug trials for treatment of MCI associated with Alzheimer's disease have undergone failure that led to development of drugs that could avert the progression of disease. However, some novel non-drug based therapies like ultrasound ablation of amyloid plaques have influenced researchers to explore the non pharmacological modalities for treatment of mild cognitive impairment. To compensate for neurodegenerative loss resulting in coexisting psychiatric disorders, neurofeedback therapy has also proven to improve behavioral outcomes through inducing neuroplasticity. The aim of the current review is to highlight the pathophysiological aspects of mild cognitive impairment leading to dementia that could be addressed with no pharmacological interventions and to understand the mechanisms behind the effects of these interventions.

## INTRODUCTION

Mild cognitive impairment (MCI) is a condition characterized by a noticeable decline in cognitive function, such as memory, attention, language, or problem-solving, that is greater than the changes typically associated with normal aging. It's important to note that not all cases of MCI are due to Alzheimer's disease. MCI can be caused by various factors, including vascular disease, Parkinson's disease, frontotemporal dementia, and other neurodegenerative conditions. Additionally, MCI can also be caused by reversible factors such as medication side effects, depression, or sleep disorders, which can be addressed with appropriate interventions. The prevalence of mild cognitive impairment ranges from 5.1% - 41% varying depending up on the regions worldwide, while its incidence ranges from 22-76.8 per 1000 persons (Pais et al., 2020). Histological findings have documented presence of amyloid beta plaques and neurofibrillary tangles as a well-established pathological mechanism underlying cognitive impairment (Iacono et al., 2014), however, with the advent of time, advanced research in the field of cognitive impairment has led to identification electrophysiological changes too. This review aims to discuss precisely the electrophysiological basis of Alzheimer's disease and neurophysiological treatment strategies for improvement of symptoms in mild cognitive impairment associated with AD.

1. **Electrophysiological findings in dementia:**

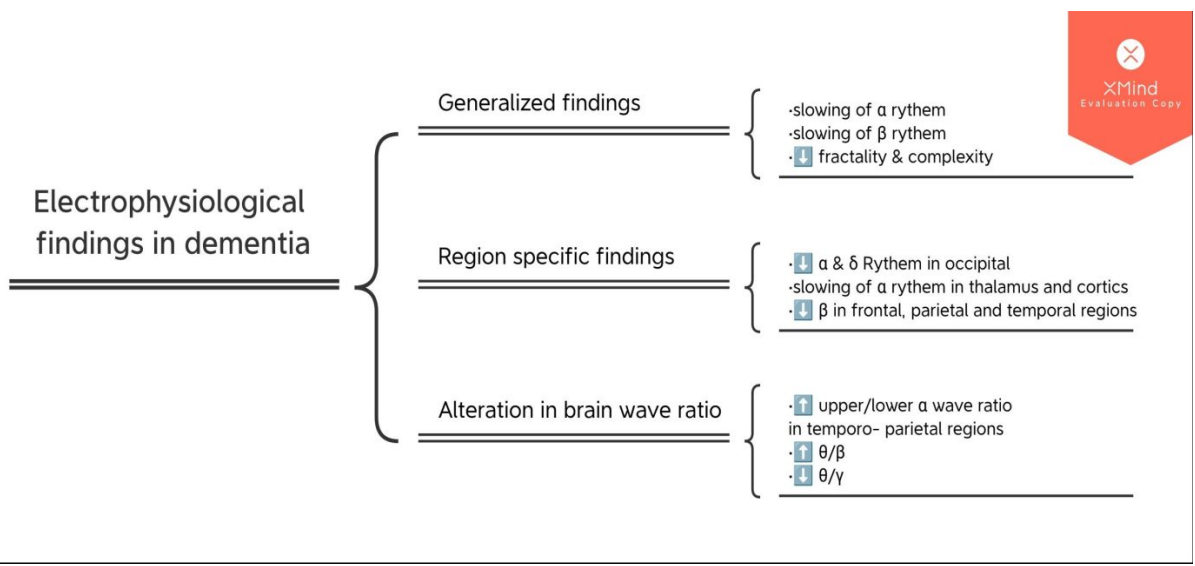

*Fig 1. Mind map – Electrophysiological findings in dementia*

## 1.1. Generalized electrophysiological findings:

Generalized slowing of brain waves has been observed in dementia (Hamilton et al., 2021). While this finding is non-specific, owing to age as a confounding factor, distinctive slowing patterns have been identified in pathological Alzheimer's dementia as compared to ageing. While generalized slowing of alpha rhythm beyond 8 Hz has been documented in organic brain syndromes (Garces et al., 2013), loss of alpha rhythm in dementia has been attributed to Alzheimer's disease. Similarly beta activity is also reduced in the brain while slow rhythms like theta and delta are dominant (Rodriguez et al., 2011). Fractality and complexity in brain signals defines the dynamic working of the brain (Herman et al., 2009; Nicolae & Ivanovici, 2021). The fractality and complexity are postulated to be reduced in neurodegenerative disorders (Zueva, 2015). While atypical changes in temporal-scale specific fractal patterns have also been identified in Alzheimer's disease (Nobukawa et al., 2019). Reduced complexity and Higuchi's fractal dimensions have also been identified in the electroencephalograms of AD patients (Besthorn et al., 1995; Smits et al., 2016). Distortion of sensory information in the spatial and temporal dimensions occur in neurodegenerative disorders like mild cognitive impairment which result in reduced fractal complexity in patterns generated by sensory neurons (Zueva, 2015). In normal ageing, fractal complexity is reduced in the right hemisphere more than the left hemisphere, while a distinct reduction in fractal complexity in Alzheimer's disease is evident in temporal and occipital regions (Smits et al., 2016), a finding that distinguishes electrophysiological processes in normal and pathological conditions. Since loss of cognitive skills mainly results from pathophysiological processes in cortex, biochemical changes in cortical neurons intricately associate with electrophysiological changes that ultimately result in symptoms of cognitive impairment.

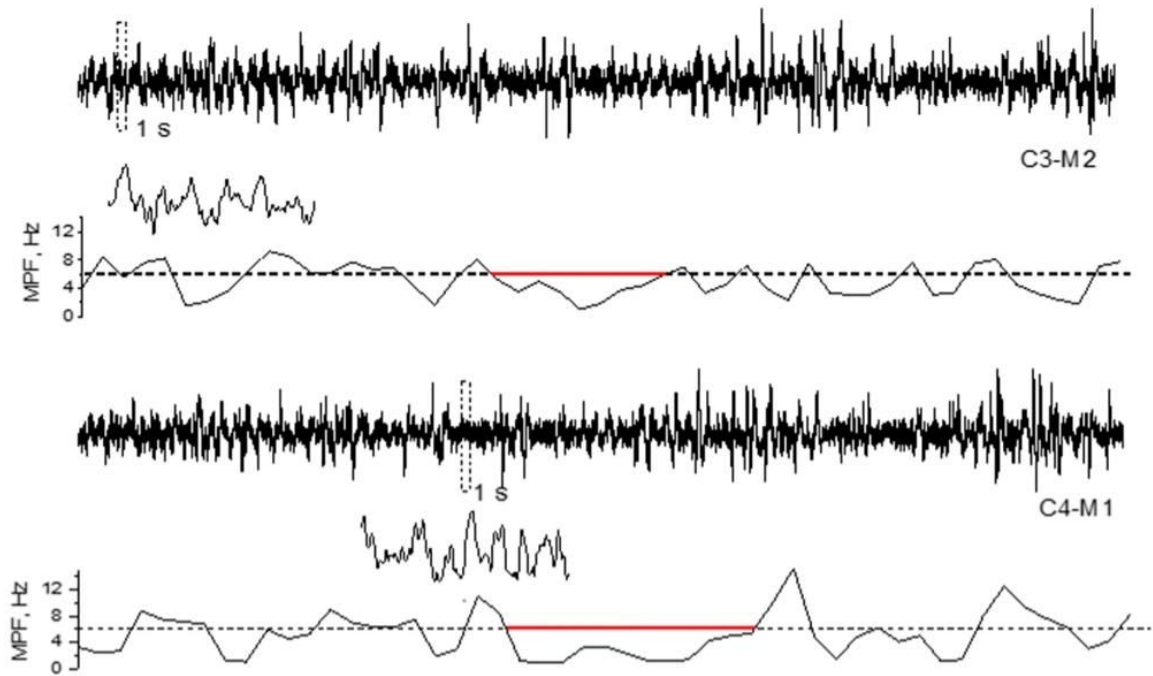

*Fig 2. EEG of patients suffering from MCI showing Paroxysmal slow wave events. (M Li et al., 2022)*

## 1.2. Region Specific brain waves:

A decline in alpha rhythm in the occipital region has been observed along with an increased magnitude of occipital delta (Babiloni et al., 2004). A computational Thalami-cortical-thalamic model has also revealed slowing of alpha rhythms in Alzheimer's Disease (Li et al., 2020). It has been reported that MCI and AD patients show a significant decreased in parietal sleep spindle densities (Gorgoni et al., 2016). Decrease beta wave coherence has also been reported in frontal, parietal and temporal regions in response higher cognitive and motor tasks in Alzheimer's disease (Radinskaia & Radinski, 2022).

### 1.3. Alterations in brain wave ratio:

Alpha/theta ratio, alpha1/theta ration and alpha2/theta ratio have been described as a promising marker for detection of early onset Alzheimer's disease (Ozbek et al., 2021). Upper/lower alpha wave ratio is increased in temporo-parietal regions in AD (Moretti et al., 2013). The umbrella of alpha/theta ratios has shown significant discrimination among frontal, parietal, central, occipital and temporal lobes in early and late onset Alzheimer's disease and also with healthy controls (Ozbek et al., 2021; Schmidt et al., 2013). An eye open theta/beta ratio has also been shown as a strong discriminator of Alzheimer's disease from healthy controls (Miao et al., 2021). Theta/beta ratios have been shown to be increased in ADHD, however, direct relationship of theta/beta ration with Alzheimer's disease is still under consideration (Picken et al., 2020). It is known that ADHD shares symptoms with Alzheimer's disease, it is postulated that theta/beta ratio also hold a great value in diagnosing Alzheimer's disease (Al-Nuaimi et al., 2021). Theta-gamma coupling, which is a crucial process behind working memory, is significantly reduced in Alzheimer's disease (Goodman et al., 2018).

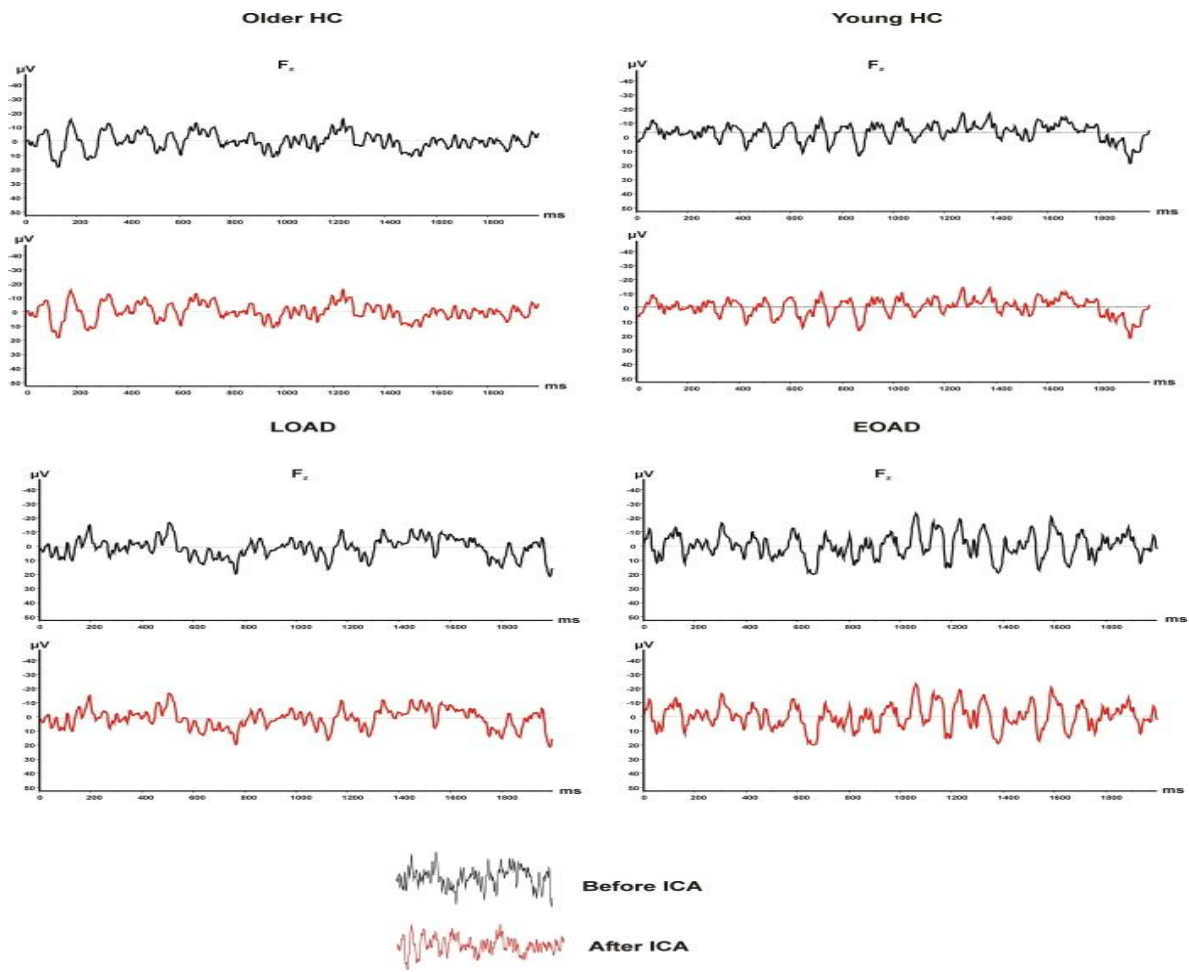

*Fig 3. Resting-state EEG alpha/theta power ratio discriminates early-onset Alzheimer's disease from healthy controls (Ozbek et al., 2021)*

1.4. **Biochemical basis of electrophysiological deficits in mild cognitive impairment** At the cellular level, generalized decrease in resting potential and increase in input resistance has been reported in both early and advanced stages of tauopathy in frontal cortical pyramidal neurons (Crimins et al., 2012). Cortical pyramidal cells are known to play a crucial role in learning and memory (Rolls, 2021), while dendritic spines in pyramidal neuron are responsible for propagation of action potential resulting in formation, processing and storage of memory (Araya, 2014). Intraneuronal tauopathy in cortical pyramidal cells results in loss of dendritic spines and thus contribute to pathological processes and symptoms in mild cognitive impairment (Merino-Serrais et al., 2013). Neurodegenerative disorders involve tauopathy as the most crucial pathological process, however, it is not limited to neurons only, but also affect glial cells (Kovacs, 2017). Tauopathies in glial cells results in neuroinflammation that further accelerates neurodegeneration (Leyns & Holtzman, 2017), moreover, it also results in functional deficits in brain (Kahlson & Colodner, 2015). The cerebellar cortex on the other hand has Purkinje cells as the only output neurons, responsible for transmitting inhibitory signals in the deep cerebellar nuclei that produce two types of action potentials that are, simple spikes and complex spikes (Hausser & Clark, 1997). These spikes contribute to a phenomenon termed 'gain modulation' (Azouz, 2005) and is required to adjust to the changing environment. Gain modulation controls the spatiotemporal synaptic integration (Palmer et al., 2010) required for sensorimotor regulation. Complex spikes are required to initiate long-term potentiation (LTP) while simple spikes provide inhibitory responses to facilitate long-term depression (LTD) (Miall et al., 1998; Zempolich et al., 2021) . Increase in simple spikes in the cerebellar region has also been reported in Alzheimer's disease correlated to increased β-secretase BACE1 activity resulting amyloid plaque formation (Cheron et al., 2022).

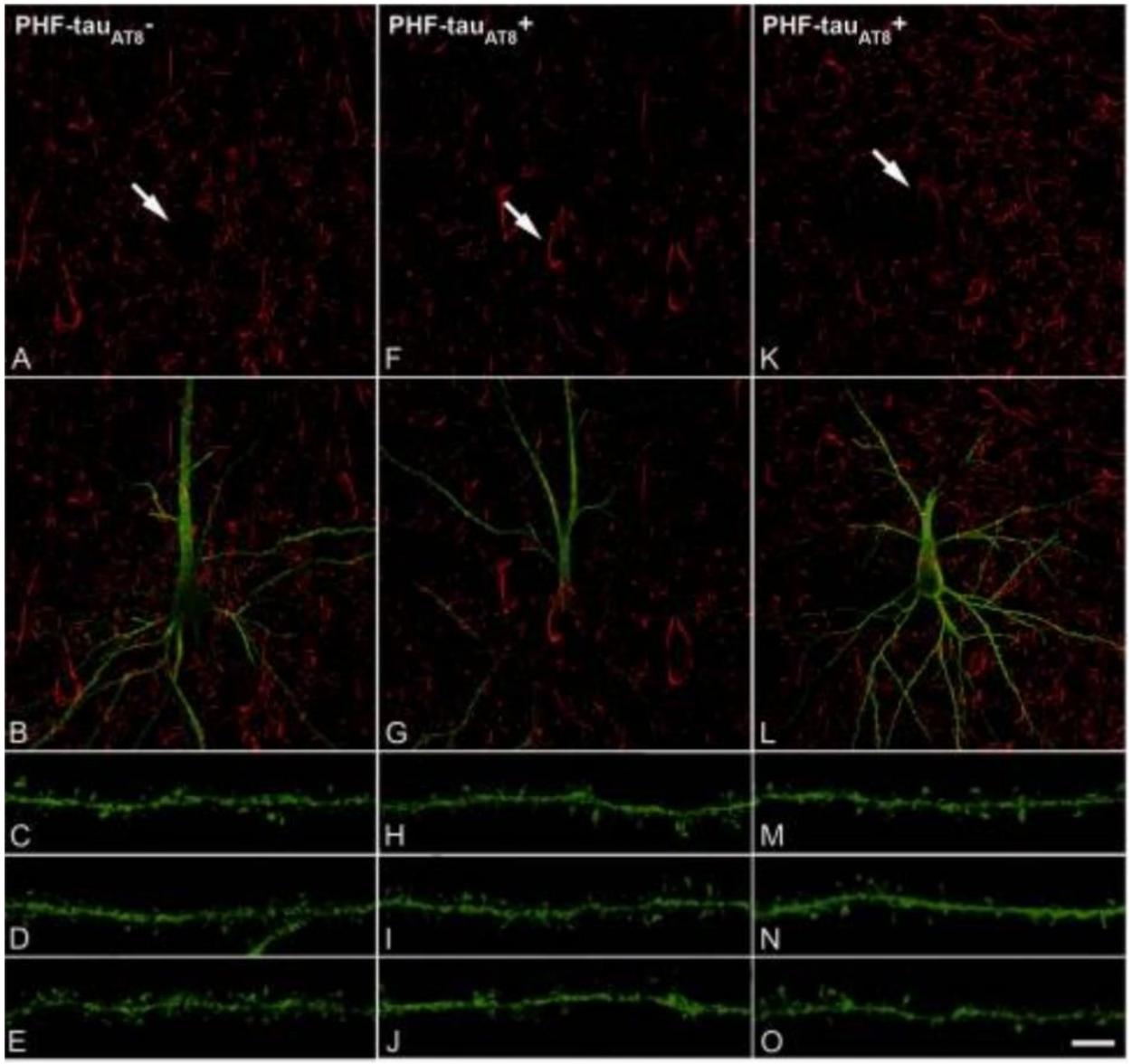

*Fig 4. Immunohistochemical staining showing tau deposits in neurons (Merino-Serrais et al., 2013)*

## 2. Modalities of treatment

### 2.1. Pharmacological treatment of Alzheimer's dementia

The biochemical modalities of treatment have ranged from chemical compounds to gene therapies recently. The cholinergic hypothesis led to the development of cholinesterase inhibitors like rivastigmine that now remains a drug of choice for treating Alzheimer's dementia. Galantamine, memantine and donepezil are also emerging modes of drug based treatment in Alzheimer's' disease. Gene therapies for nerve growth factor (NGF), Brain derived neurotrophic factor (BDNF), amyloid precursor protein (APP), Presenilin 1 and 2 (PSEN1 and PSEN2) protein, BACE 1 and ApoE are also under rigorous research (Se Thoe et al., 2021). Immunotherapies for amyloid beta are also under clinical trials, however, they haven't yet been approved. The drug based treatment presents with the adverse effects associated with hepatic and gastro-intestinal function renders the drug non-favorable. Moreover, reducing the dosage potency results in loss of efficacy of these drugs. Therefore, the treatment approaches are currently shifting toward non-drug based modalities that we discuss in the following.

### 2.2. Non-Pharmacological Treatment strategies for Alzheimer's dementia

Non-drug based treatments for Alzheimer's dementia originate from the fact that Alzheimer's disease has also been classified as type III diabetes, which indicates its metabolic origin. Life style modifications therefore, have been deemed useful in reducing risk of dementia (Krell-Roesch et al., 2018; Lautenschlager et al., 2010). Cognitive interventions have been recently explored for their effectiveness in improving Alzheimer's symptoms. These range from cognitive stimulation (CST), cognitive training and cognitive rehabilitation (D'Onofrio G et al., 2016; Zucchella et al., 2018).

### 2.2.1. Cognitive stimulation Treatment:

Variable reports have been documented for the effectiveness of cognitive stimulation for treating dementia, where the type of CST was the most significant predictor of its effectiveness (Holden et al., 2021). Dose and duration of CST, individual capability of the CST facilitators, supervised training and service, professional degree and outcome monitoring strongly affected the effectiveness of CST. Yet, moderate benefits of CST, with improvement in quality of life have been reported in Alzheimer's disease (Kim et al., 2017), holding less value for the caregivers of dementia patients (Aguirre et al., 2014). The recommended number of session for effective cognitive stimulation is 14, at which substantial improvements in cognition, anxiety and behavioral symptoms can be observed (Paddick et al., 2017). Therapeutic regimen usually presents with multiple sessions held for a longer time duration. While short therapies have shown improvement in various other disorders, cognitive impairment in dementia requires longer time duration of CST to exhibit its useful effects (Chen et al., 2019). Longer time durations can be justified for disorders that hold neurodegeneration as a the underlying pathological mechanism because the lost neurons carrying important information needs to be replaced by healthy neurons forming new connections with old information, the reason why CST has shown greater improvement in combination with acetylcholinesterase inhibitors (Chen et al., 2019). The effect of combined cognitive and physical training has also been studied where reduction in beta, theta and delta rhythms have been recorded. These changes were mainly observed in the precuneus region (Styliadis et al., 2015). Precuneus is a crucial part in the default mode network, which comprises of connections between precuneus, posterior cingulate cortex (PCC), inferior parietal cortex, medial temporal lobes, medial frontal cortex, and anterior cingulate cortex (Raichle & Snyder, 2007) and is known to be vulnerable to atrophy in mild cognitive impairment (Buckner et al., 2005). Improvement in beta, theta and the delta rhythms after cognitive training therefore signify the effects of cognitive trainings on brain function. Similarly, repetitive transcranial magnetic stimulation has shown increase in delta activity at the temporal region while a nonsignificant increase of the log EEG power for alpha band over the frontal and temporal regions, beta band over the frontal region, theta band over the frontal, temporal, and parieto-occipital regions, and delta band over the frontal and parieto-occipital regions (Gandelman-Marton et al., 2017). For long lasting effects of cognitive stimulation, maintenance therapy has also been devised (Aguirre et al., 2010), however, compliance to report to the service centers has played a great role in limiting the dose, duration and therefore, effectiveness of CST (Holden et al., 2021)

### 2.2.2. Psychotherapy

Psychotherapeutic approaches deploying psychodynamic therapy, interpersonal therapy, supportive counseling (Orgeta et al., 2014) have manifested improvement in depression and anxiety linked to mild cognitive impairment. The efficacy of psychotherapy in the treatment of mild cognitive impairment requires rigorous consideration of various factors like content, repetition, implementation of external memory aids and inclusion of caregivers into therapeutic process (Linnemann & Fellgiebel, 2017). Qualitative analysis in the systematic reviews have uncovered the beneficial effects of psychotherapy on cognitive restructuring, disease-state acceptance, and quality of life (Rostamzadeh et al., 2022). Interpersonal therapy holds value as maintenance therapy that needs to be combined with other therapies or medication to exert its effects (Miller & Reynolds, 2007). Interpersonal therapy requires integration of the caregiver in the therapeutic process. The techniques utilized in interpersonal therapy include role conflict as an important approach to integrate the care giver in the process of therapy, thereby reducing the sense of isolation in the patients with cognitive dysfunction (Miller & Reynolds, 2007). Group therapy on the other hand also serves to alleviate the state of helplessness and improve acceptance about the disease condition and memory related problems (Joosten-Weyn Banningh et al., 2011). Also these patients have been shown to cope up with the future uncertainty about the progression of disease and their overall state of health (Joosten-Weyn Banningh et al., 2011). Social support intervention has been deemed useful in reducing depression and improving quality of life in early-stage memory loss. These participants have also reported a higher self-esteem, yet these studies present with limitations in the design and inclusion of participants (Leung et al., 2015). Therefore, social support interventions are still under consideration for the clinical recommAwendation.

### 2.2.3. Aroma therapy

Pre-clinical trials on aroma therapy have demonstrated improvement in cognitive function via reduction in amyloid beta and phosphorylated tau with increase in brain-derived neurotrophic factor (BDNF) in the olfactory bulb (Okuda et al., 2020). Aromas from lavender, lemon, orange and cedar extracts, either alone, or in combination have been tested for their efficacy in improving agitate, behavioral and psychological symptoms and cognition, but the results obtained do not provide any clinical significance for their use in managing these behaviors. Aromatherapy has been found to be useful in the treatment of dementia (Forrester et al., 2014), yet, memory and executive functioning needs to be addressed in these studies for assessing effects of aromatherapy (Ball et al., 2020).

### 2.2.4. Art Therapy

Art therapy has been utilized by many researchers for its effects on patients suffering from mild cognitive impairment (Chancellor et al., 2014; de Souza et al., 2022). Mahendran, R., *et al*. have reported improvement in depression and anxiety related to mild cognitive impairment along with improvement in telomere length, however lack of statistical significance renders these findings less useful clinically (Mahendran et al., 2018). Expressive art therapy on the other hand has shown improvement in many other function in addition to anxiety and depression that include social relationships and psychological domains, language and cognitive functions (Yan et al., 2021).

### 2.2.5. Music Therapy

Music therapy has been widely implied for therapy in a number of psychological conditions. Its effects on improving cognitive functions have been found to be positive (Fang et al., 2017; Lyu et al., 2018; Sherratt et al., 2004). Since music provides an auditory stimulus, auditory evoked potentials are therefore generated, which have been found beneficial in disorders of consciousness (Magee et al., 2015). The pathological mechanisms underlying disorders of consciousness include oxygen deprivation, diffuse axonal injury, loss of facilitation and withdrawal of excitatory activity amongst other mechanisms (Edlow et al., 2021), these mechanisms are also shared in traumatic encephalopathy and Alzheimer's disease (Katsumoto et al., 2019). Therefore, auditory stimulus like music tends to show improvement in cognitive functions (Moreno-Morales et al., 2020). The mode of delivery of music has also been evaluated in this regard and it is found that musical stimuli using digital technologies are also of great value (Han et al., 2020) in improving cognitive functions. Moreover, therapeutic effects of music have also been evident in different types of higher cognitive functions like general cognition, executive function and episodic memory (Ito et al., 2022).

### 2.2.6. Technology aided therapies

#### *2.2.6.1. Ultrasound*

Technological advancements have led to more technologically oriented treatment strategies like the use of ultrasounds (Gotz et al., 2021; Nicodemus et al., 2019). Preclinical trials on focused ultrasounds have found significant reduction in beta amyloid plaques (Hsu et al., 2018), along with significant inhibition of GSK-3 protein, which contributes to phosphorylation of tau protein in its active state. Pre-clinical trials on low intensity pulsed ultrasound has shown significant improvement in cognitive functions via activation of endothelial nitric oxide synthase (Eguchi et al., 2018). Scanning ultrasound has also shown improvement in spatial memory, and long-term potentiation (LTP). Scanning ultrasound along with micro bubbles injected for enhancing blood brain barrier permeability has shown improvement in LTP while clearing amyloid plaques and tau tangles resulting in improved memory in transgenic mice(Gotz et al., 2021). Alongside cognitive improvement, focused ultrasound has also shown improvement in fine and gross motor scores (Nicodemus et al., 2019), while transcranial pulsed ultrasound has shown also shown improvement in cortical thickness (Popescu et al., 2021).

#### *2.2.6.2. Photobiomodulation*

Pre-clinical trials of photobiomodulation on transgenic Alzheimer's disease rat models have shown reduction in neuroinflammation, improvement in regulation of glial cells polarization, preservation of mitochondrial dynamics, suppression of oxidative damage and clearance of amyloid plaques (Yang et al., 2022). Linda L Chao has also shown increased cerebral perfusion, increased connectivity between posterior cingulate cortex and lateral parietal nodes in the default mode network after Photobiomodulation (Chao, 2019). Transcranial pulsed intranasal photobiomodulation utilizing 810 nm, 10 Hz pulsed light-emitting diode device has returned significant results in clinical trials, exhibiting improved cognitive functions, reduced anxiety, anger outburst and wandering along with improvement in sleep (Saltmarche et al., 2017). While meta-analysis on the effect of photobiomodulation in improving cognitive functions in dementia have exhibited significantly positive results. Clinical trials with larger sample sizes have been recommended to deduce its effectiveness as an approved modality for treatment (Salehpour et al., 2021).

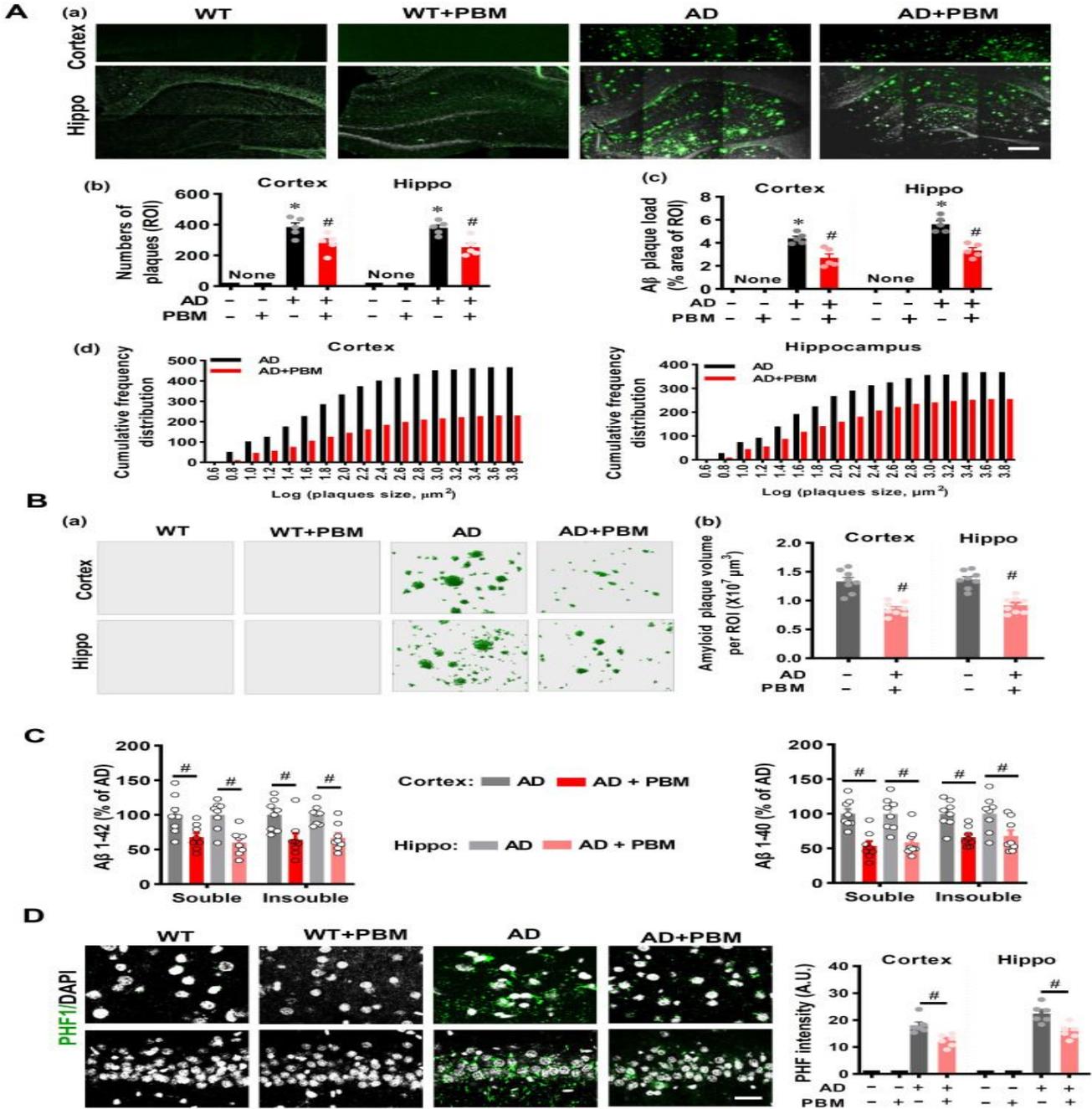

*Fig 5. PBM showing reduction in Amyloid deposits (AD) Yang et al., 2022*

### 2.2.6.3. Virtual reality:

Virtual reality and gaming also emerged as a useful tool in improving cognitive functions in Alzheimer's disease like attention (Kalova et al., 2005), executive functions (Kalova et al., 2005; Sauzeon et al., 2016), and memory (Abichou et al., 2017). Most studies have demonstrated the use of virtual reality for definitive diagnosis of dementia and Alzheimer's' disease (Clay et al., 2020). The therapeutic purposes have explored its usefulness in improving the emotional health in dementia patients (Appel et al., 2021). Negative emotions have been successfully regulated using a virtual reality zoo, where Alzheimer's disease patients have been made to interact with different animals (Ben Abdessalem et al., 2021). Virtual zoo therapy utilized real-time gesture recognition system that predicted interaction of these patients with virtual animals (Ben Abdessalem et al., 2021). Higher sense of satisfaction has also been reported in nursing homes utilizing virtual reality as an added therapeutic modality (Kim et al., 2021). Semi immersive virtual reality technology has been found to hold greater effect than full immersive technology (Kim et al., 2019) in dementia and mild cognitive impairment. Virtual reality exploiting instrumental activities of daily living as a therapeutic modality for dementia patients has returned positive results in improvement of global cognition (Oliveira et al., 2021) and has been suggested as an effective tool for cognitive stimulation therapies too.

## 2.2.7. Neurofeedback

Neurofeedback relies on the principle of operant conditioning. As loss of cognitive functions occur in Alzheimer's disease, operant conditioning using neurofeedback has shown to improve cognitive symptoms in the disease in preliminary studies (Arns et al., 2017; Luijmes et al., 2016). Attention training using neurofeedback has been performed as a prevention strategy for Alzheimer's disease in ageing brains (Jiang et al., 2017). Working memory has also exhibited improvement using neurofeedback protocols (Jiang et al., 2022). While task-based neurofeedback training has shown improvement in executive functioning of healthy adults (Berman; Hosseini et al., 2016), effects of neurofeedback training remains an area of interest with reference to executive functioning.

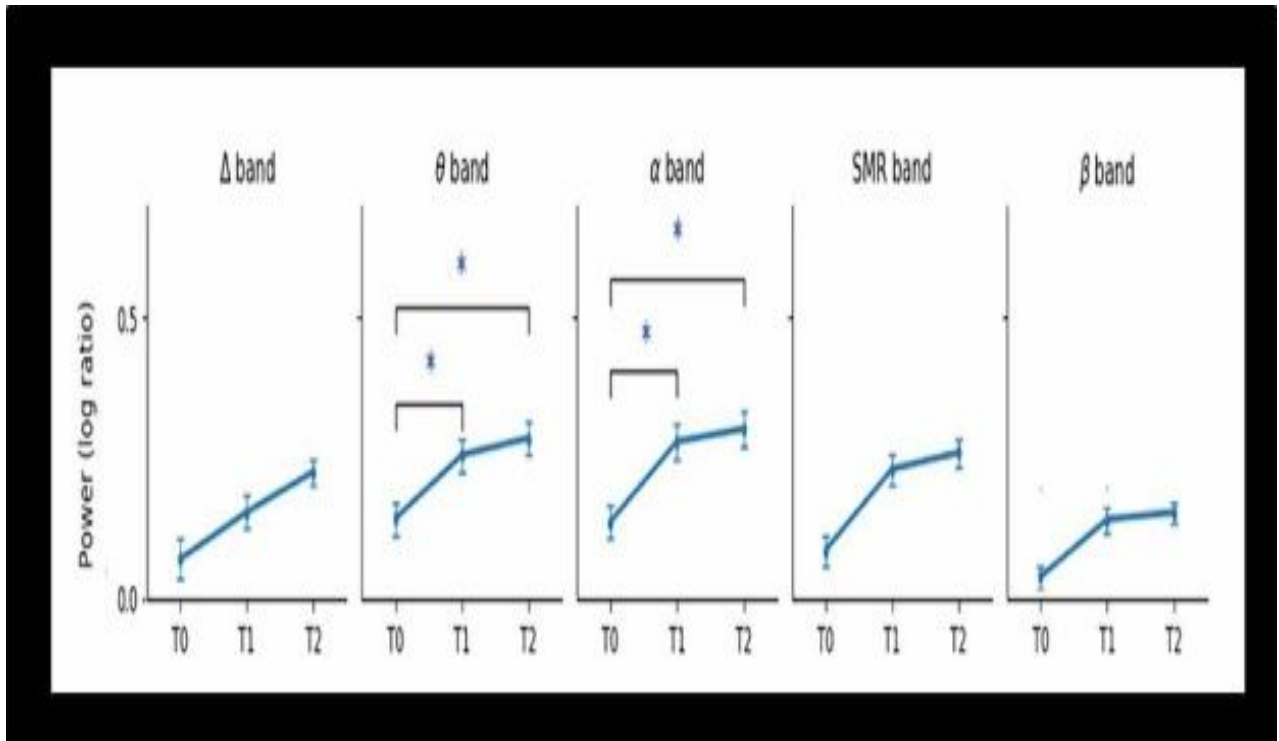

*Fig 6. Changes in power of EEG bands after neurofeedback mechanism Marlats et al., 2020*

### 2.2.7.1.1. *Regulation of alpha beta bands:*

Down regulation of alpha band or up-regulation of beta band has been successfully implied in the neurofeedback at the central electrodes using motor-imagery at the central electrodes in mild cognitive impairment (Mendoza Laiz et al., 2018).While increase in alpha power at central parietal region resulted in composite memory improvement (Lavy et al., 2019). Enhancement of sensory-motor rhythm (SMR) band while reducing beta and theta rhythms in CZ region also exhibited improvement in Goldberg Anxiety 322 Scale (GAS) and the Wechsler Adult Intelligence Score IV (Marlats et al., 2020). Up-regulation of beta/alpha ratio has also been reported to enhance spatial working memory, visual information processing and Cambridge 334 Neuropsychological Test Automated Battery (CANTAB) scores (Coull et al., 1995). Increasing upper alpha has shown improvement in cognitive performance.

### 2.2.7.2. *Alpha theta protocol*

Theta band has been shown to facilitate memory consolidation, hence alpha theta protocol has been tested for its effectiveness in working memory. Up-regulation of theta has been shown to improve alpha self-regulation and enhance working memory (Reis et al., 2016). Enhancing upper alpha has also shown improvement in memory (Lavy et al., 2021).

### 2.2.7.3. *Beta alpha protocol*

Decreasing alpha and increasing beta at the central region has also been shown to improve cognitive symptoms (Lavy et al., 2021). Enhancing beta/alpha ratio at AF3 and AF4 region using a game-based neurofeedback has shown improvement in rapid visual processing and spatial working memory (Jirayucharoensak et al., 2019). Enhancing beta power has also shown improvement in cognitive functions in mild cognitive impairment (Jang et al., 2019). Recent studies have proposed SMR/theta and beta/theta protocols for improving cognitive outcomes in mild cognitive impairment (Marlats et al., 2019).

### 2.2.8. Physiotherapy

### *2.2.8.1. Aerobic therapy*

Individuals with Alzheimer's disease experienced improvements in gait, balance, and cognitive impairment after undergoing one month of low-frequency, short-duration, practical physical therapy aimed at addressing motor impairment and functional limitations (Andersen et al., 2014). The use of acetylcholinesterase inhibitors have been reported to affect physical outcomes in individuals with cognitive impairment (Longhurst et al., 2020). Inactivity and unhealthy habits are closely associated with the development of cognitive decline (Marques et al., 2019).

### *2.2.8.2. Hydrotherapy*

Amyloid is deposited in the brain as a consequence of blood flow; during blood flow, amyloid is transferred from the blood stream into the brain resulting in amyloid deposition. Decrease in amyloid deposits lead to improved in symptoms of Alzheimer's disease (JH Kim et al., 2018). Aquatic therapy promote neuroplasticity by providing a unique sensory environment that can enhance sensory-motor integration and proprioceptive feedback (Lambeck S. Et al., 2020).

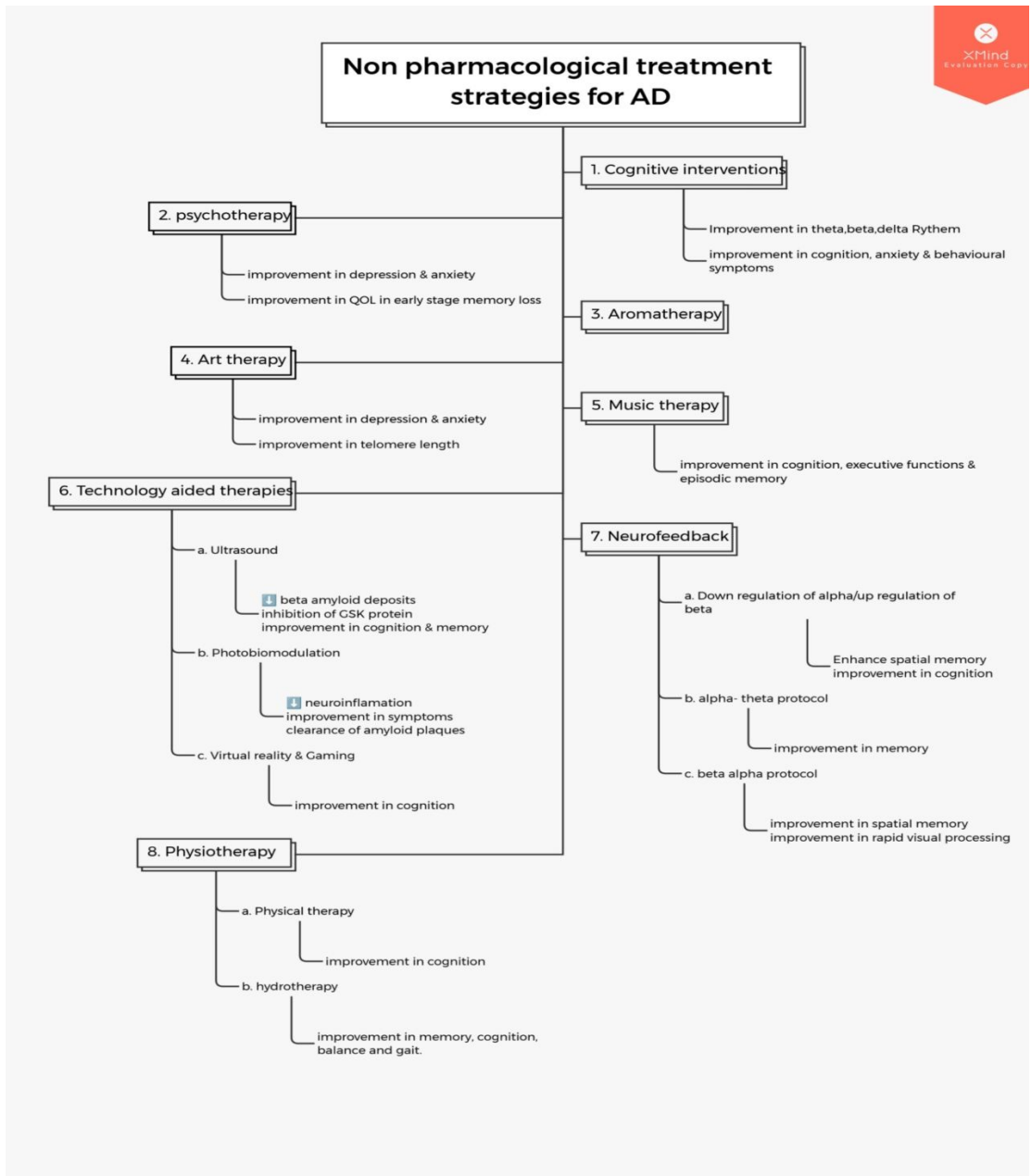

*Fig 7. Flowchart – Non pharmacological treatment strategies for AD*

## 3. Conclusion.

Mild cognitive impairment is challenging in its diagnosis and treatment due to remarkable similarities with conventional ageing. Understanding of pathological mechanisms led to development of pharmaceutical treatment options, however, the cost and adverse effects of this intervention led to the need for the development of alternative and cost-effective approaches. Identification of electrophysiological changes revealed a novel approach for treatment of mild cognitive impairment, thereby, widening the arena for exploring alternative therapies. Psychological treatments like cognitive stimulation and psychotherapy have returned some clinically useful results in improving symptoms of dementia and also reported better quality of life outcomes. Sensory stimulation via aroma therapy, art therapy and photobiomodulation have reported some useful results, yet, there is still deficiency of proper evidence and more studies are needed to explore highly effective methods for improvement in high order functioning in cognitive impairment. Technologically aided therapies like ultrasound, first made its way in exploring this mode of therapy with least side effects. Use of virtual reality and gaming came later and has introduced a new mode of therapy for treating symptoms of dementia, however, specialized protocol for multiple executive functions, recall functions, and motor coordination needs to be developed to assess the range of therapeutic effects that it can provide. Alterations in electrophysiological parameters like generalized slowing of alpha wave, alterations theta/beta and alpha/theta ratio led to development of brain wave specific protocols for improving cognitive outcomes. The therapies developed so far offer a wide range of benefits with some limitations, yet, a smart mode incorporating multiple effects on brain function still needs to be developed for a concise mode of alternative treatment in dementia to improve cognitive outcomes.

## 4. References

Abichou, K., La Corte, V., & Piolino, P. (2017). Does virtual reality have a future for the study of episodic memory in aging? *Geriatr Psychol Neuropsychiatr Vieil*, *15*(1), 65-74. https://doi.org/10.1684/pnv.2016.0648 (La realite virtuelle a-t-elle un avenir pour l'etude de la memoire episodique dans le vieillissement ?)

Aguirre, E., Hoare, Z., Spector, A., Woods, R. T., & Orrell, M. (2014). The effects of a Cognitive Stimulation Therapy [CST] programme for people with dementia on family caregivers' health. *BMC Geriatr*, *14*, 31. https://doi.org/10.1186/1471-2318-14-31

Aguirre, E., Spector, A., Hoe, J., Russell, I. T., Knapp, M., Woods, R. T., & Orrell, M. (2010). Maintenance Cognitive Stimulation Therapy (CST) for dementia: a single-blind, multi-centre, randomized controlled trial of Maintenance CST vs. CST for dementia. *Trials*, *11*, 46. https://doi.org/10.1186/1745-6215-11-46

Al-Nuaimi, A. H., Bluma, M., Al-Juboori, S. S., Eke, C. S., Jammeh, E., Sun, L., & Ifeachor, E. (2021). Robust EEG Based Biomarkers to Detect Alzheimer's Disease. *Brain Sci*, *11*(8). https://doi.org/10.3390/brainsci11081026

Appel, L., Ali, S., Narag, T., Mozeson, K., Pasat, Z., Orchanian-Cheff, A., & Campos, J. L. (2021). Virtual reality to promote wellbeing in persons with dementia: A scoping review. *J Rehabil Assist Technol Eng*, *8*, 20556683211053952. https://doi.org/10.1177/20556683211053952

Araya, R. (2014). Input transformation by dendritic spines of pyramidal neurons. *Front Neuroanat*, *8*, 141. https://doi.org/10.3389/fnana.2014.00141

Arns, M., Batail, J. M., Bioulac, S., Congedo, M., Daudet, C., Drapier, D., Fovet, T., Jardri, R., Le-Van-Quyen, M., Lotte, F., Mehler, D., Micoulaud-Franchi, J. A., Purper-Ouakil, D., Vialatte, F., & group, N. E. (2017). Neurofeedback: One of today's techniques in psychiatry? *Encephale*, *43*(2), 135-145. https://doi.org/10.1016/j.encep.2016.11.003

Azouz, R. (2005). Dynamic spatiotemporal synaptic integration in cortical neurons: neuronal gain, revisited. *J Neurophysiol*, *94*(4), 2785-2796. https://doi.org/10.1152/jn.00542.2005

Babiloni, C., Binetti, G., Cassetta, E., Cerboneschi, D., Dal Forno, G., Del Percio, C., Ferreri, F., Ferri, R., Lanuzza, B., Miniussi, C., Moretti, D. V., Nobili, F., Pascual-Marqui, R. D., Rodriguez, G., Romani, G. L., Salinari, S., Tecchio, F., Vitali, P., Zanetti, O., Zappasodi, F., & Rossini, P. M. (2004). Mapping distributed sources of cortical rhythms in mild Alzheimer's disease. A multicentric EEG study. *Neuroimage*, *22*(1), 57-67. https://doi.org/10.1016/j.neuroimage.2003.09.028

Ball, E. L., Owen-Booth, B., Gray, A., Shenkin, S. D., Hewitt, J., & McCleery, J. (2020). Aromatherapy for dementia. *Cochrane Database Syst Rev*, *8*, CD003150. https://doi.org/10.1002/14651858.CD003150.pub3

Ben Abdessalem, H., Ai, Y., Marulasidda Swamy, K. S., & Frasson, C. (2021). Virtual Reality Zoo Therapy for Alzheimer's Disease Using Real-Time Gesture Recognition. *Adv Exp Med Biol*, *1338*, 97-105. https://doi.org/10.1007/978-3-030-78775-2_12

Berman, M., & Frederick, J. Efficacy of neurofeedback for executive and memory function in dementia. (unpublished manuscript). https://www.bmedreport.com/archives/5778

Besthorn, C., Sattel, H., Geiger-Kabisch, C., Zerfass, R., & Forstl, H. (1995). Parameters of EEG dimensional complexity in Alzheimer's disease. *Electroencephalogr Clin Neurophysiol*, *95*(2), 84-89. https://doi.org/10.1016/0013-4694(95)00050-9

Buckner, R. L., Snyder, A. Z., Shannon, B. J., LaRossa, G., Sachs, R., Fotenos, A. F., Sheline, Y. I., Klunk, W. E., Mathis, C. A., Morris, J. C., & Mintun, M. A. (2005). Molecular, structural, and functional characterization of Alzheimer's disease: evidence for a relationship between default activity, amyloid, and memory. *J Neurosci*, *25*(34), 7709-7717. https://doi.org/10.1523/JNEUROSCI.2177-05.2005

Chancellor, B., Duncan, A., & Chatterjee, A. (2014). Art therapy for Alzheimer's disease and other dementias. *J Alzheimers Dis*, *39*(1), 1-11. https://doi.org/10.3233/JAD-131295

Chao, L. L. (2019). Effects of Home Photobiomodulation Treatments on Cognitive and Behavioral Function, Cerebral Perfusion, and Resting-State Functional Connectivity in Patients with Dementia: A Pilot Trial. *Photobiomodul Photomed Laser Surg*, *37*(3), 133-141. https://doi.org/10.1089/photob.2018.4555

Chen, J., Duan, Y., Li, H., Lu, L., Liu, J., & Tang, C. (2019). Different durations of cognitive stimulation therapy for Alzheimer's disease: a systematic review and meta-analysis. *Clin Interv Aging*, *14*, 1243-1254. https://doi.org/10.2147/CIA.S210062

Cheron, G., Ristori, D., Marquez-Ruiz, J., Cebolla, A. M., & Ris, L. (2022). Electrophysiological alterations of the Purkinje cells and deep cerebellar neurons in a mouse model of Alzheimer disease (electrophysiology on cerebellum of AD mice). *Eur J Neurosci*. https://doi.org/10.1111/ejn.15621

Clay, F., Howett, D., FitzGerald, J., Fletcher, P., Chan, D., & Price, A. (2020). Use of Immersive Virtual Reality in the Assessment and Treatment of Alzheimer's Disease: A Systematic Review. *J Alzheimers Dis*, *75*(1), 23-43. https://doi.org/10.3233/JAD-191218

Coull, J. T., Middleton, H. C., Robbins, T. W., & Sahakian, B. J. (1995). Contrasting effects of clonidine and diazepam on tests of working memory and planning. *Psychopharmacology (Berl)*, *120*(3), 311-321. https://doi.org/10.1007/BF02311179

Crimins, J. L., Rocher, A. B., & Luebke, J. I. (2012). Electrophysiological changes precede morphological changes to frontal cortical pyramidal neurons in the rTg4510 mouse model of progressive tauopathy. *Acta Neuropathol*, *124*(6), 777-795. https://doi.org/10.1007/s00401-012-1038-9

D'Onofrio G, Sancarlo D, Seripa D, Ricciardi F, Giuliani F, Panza F, & al., e. (2016). Chapter 18: non-pharmacological approaches in the treatment of dementia. Update on dementia. In D. Moretti (Ed.), *Update on Dementia. Intech* (pp. 477–491).

de Souza, L. B. R., Gomes, Y. C., & de Moraes, M. G. G. (2022). The impacts of visual Art Therapy for elderly with Neurocognitive disorder: a systematic review. *Dement Neuropsychol*, *16*(1), 8-18. https://doi.org/10.1590/1980-5764-DN-2021-0042

Edlow, B. L., Claassen, J., Schiff, N. D., & Greer, D. M. (2021). Recovery from disorders of consciousness: mechanisms, prognosis and emerging therapies. *Nat Rev Neurol*, *17*(3), 135-156. https://doi.org/10.1038/s41582-020-00428-x

Eguchi, K., Shindo, T., Ito, K., Ogata, T., Kurosawa, R., Kagaya, Y., Monma, Y., Ichijo, S., Kasukabe, S., Miyata, S., Yoshikawa, T., Yanai, K., Taki, H., Kanai, H., Osumi, N., & Shimokawa, H. (2018). Whole-brain low-intensity pulsed ultrasound therapy markedly improves cognitive dysfunctions in mouse models of dementia - Crucial roles of endothelial nitric oxide synthase. *Brain Stimul*, *11*(5), 959-973. https://doi.org/10.1016/j.brs.2018.05.012

Fang, R., Ye, S., Huangfu, J., & Calimag, D. P. (2017). Music therapy is a potential intervention for cognition of Alzheimer's Disease: a mini-review. *Transl Neurodegener*, *6*, 2. https://doi.org/10.1186/s40035-017-0073-9

Forrester, L. T., Maayan, N., Orrell, M., Spector, A. E., Buchan, L. D., & Soares-Weiser, K. (2014). Aromatherapy for dementia. *Cochrane Database Syst Rev*(2), CD003150. https://doi.org/10.1002/14651858.CD003150.pub2

Gandelman-Marton, R., Aichenbaum, S., Dobronevsky, E., Khaigrekht, M., & Rabey, J. M. (2017). Quantitative EEG After Brain Stimulation and Cognitive Training in Alzheimer Disease. *J Clin Neurophysiol*, *34*(1), 49-54. https://doi.org/10.1097/WNP.0000000000000301

Garces, P., Vicente, R., Wibral, M., Pineda-Pardo, J. A., Lopez, M. E., Aurtenetxe, S., Marcos, A., de Andres, M. E., Yus, M., Sancho, M., Maestu, F., & Fernandez, A. (2013). Brain-wide slowing of spontaneous alpha rhythms in mild cognitive impairment. *Front Aging Neurosci*, *5*, 100. https://doi.org/10.3389/fnagi.2013.00100

Goodman, M. S., Kumar, S., Zomorrodi, R., Ghazala, Z., Cheam, A. S. M., Barr, M. S., Daskalakis, Z. J., Blumberger, D. M., Fischer, C., Flint, A., Mah, L., Herrmann, N., Bowie, C. R., Mulsant, B. H., & Rajji, T. K. (2018). Theta-Gamma Coupling and Working Memory in Alzheimer's Dementia and Mild Cognitive Impairment. *Front Aging Neurosci*, *10*, 101. https://doi.org/10.3389/fnagi.2018.00101

Gorgoni, M., Lauri, G., Truglia, I., Cordone, S., Sarasso, S., Scarpelli, S., Mangiaruga, A., D'Atri, A., Tempesta, D., Ferrara, M., Marra, C., Rossini, P. M., & De Gennaro, L. (2016). Parietal Fast Sleep Spindle Density Decrease in Alzheimer's Disease and Amnesic Mild Cognitive Impairment. *Neural Plast*, *2016*, 8376108. https://doi.org/10.1155/2016/8376108

Gotz, J., Richter-Stretton, G., & Cruz, E. (2021). Therapeutic Ultrasound as a Treatment Modality for Physiological and Pathological Ageing Including Alzheimer's Disease. *Pharmaceutics*, *13*(7). https://doi.org/10.3390/pharmaceutics13071002

Hamilton, C. A., Schumacher, J., Matthews, F., Taylor, J. P., Allan, L., Barnett, N., Cromarty, R. A., Donaghy, P. C., Durcan, R., Firbank, M., Lawley, S., O'Brien, J. T., Roberts, G., & Thomas, A. J. (2021). Slowing on quantitative EEG is associated with transition to dementia in mild cognitive impairment. *Int Psychogeriatr*, *33*(12), 1321-1325. https://doi.org/10.1017/S1041610221001083

Han, E., Park, J., Kim, H., Jo, G., Do, H. K., & Lee, B. I. (2020). Cognitive Intervention with Musical Stimuli Using Digital Devices on Mild Cognitive Impairment: A Pilot Study. *Healthcare (Basel)*, *8*(1). https://doi.org/10.3390/healthcare8010045

Hausser, M., & Clark, B. A. (1997). Tonic synaptic inhibition modulates neuronal output pattern and spatiotemporal synaptic integration. *Neuron*, *19*(3), 665-678. https://doi.org/10.1016/s0896-6273(00)80379-7

Herman, P., Kocsis, L., & Eke, A. (2009). Fractal characterization of complexity in dynamic signals: application to cerebral hemodynamics. *Methods Mol Biol*, *489*, 23-40. https://doi.org/10.1007/978-1-59745-543-5_2

Holden, E., Stoner, C. R., & Spector, A. (2021). Cognitive stimulation therapy for dementia: Provision in National Health Service settings in England, Scotland and Wales. *Dementia (London)*, *20*(5), 1553-1564. https://doi.org/10.1177/1471301220954611

Hosseini, S. M. H., Pritchard-Berman, M., Sosa, N., Ceja, A., & Kesler, S. R. (2016). Task-based neurofeedback training: A novel approach toward training executive functions. *Neuroimage*, *134*, 153-159. https://doi.org/10.1016/j.neuroimage.2016.03.035

Hsu, P. H., Lin, Y. T., Chung, Y. H., Lin, K. J., Yang, L. Y., Yen, T. C., & Liu, H. L. (2018). Focused Ultrasound-Induced Blood-Brain Barrier Opening Enhances GSK-3 Inhibitor Delivery for Amyloid-Beta Plaque Reduction. *Sci Rep*, *8*(1), 12882. https://doi.org/10.1038/s41598-018-31071-8

Iacono, D., Resnick, S. M., O'Brien, R., Zonderman, A. B., An, Y., Pletnikova, O., Rudow, G., Crain, B., & Troncoso, J. C. (2014). Mild cognitive impairment and asymptomatic Alzheimer disease subjects: equivalent beta-amyloid and tau loads with divergent cognitive outcomes. *J Neuropathol Exp Neurol*, *73*(4), 295-304. https://doi.org/10.1097/NEN.0000000000000052

Ito, E., Nouchi, R., Dinet, J., Cheng, C. H., & Husebo, B. S. (2022). The Effect of Music-Based Intervention on General Cognitive and Executive Functions, and Episodic Memory in People with Mild Cognitive Impairment and Dementia: A Systematic Review and Meta-Analysis of Recent Randomized Controlled Trials. *Healthcare (Basel)*, *10*(8). https://doi.org/10.3390/healthcare10081462

Jang, J. H., Kim, J., Park, G., Kim, H., Jung, E. S., Cha, J. Y., Kim, C. Y., Kim, S., Lee, J. H., & Yoo, H. (2019). Beta wave enhancement neurofeedback improves cognitive functions in patients with mild cognitive impairment: A preliminary pilot study. *Medicine (Baltimore)*, *98*(50), e18357. https://doi.org/10.1097/MD.0000000000018357

Jiang, Y., Abiri, R., & Zhao, X. (2017). Tuning Up the Old Brain with New Tricks: Attention Training via Neurofeedback. *Front Aging Neurosci*, *9*, 52. https://doi.org/10.3389/fnagi.2017.00052

Jiang, Y., Jessee, W., Hoyng, S., Borhani, S., Liu, Z., Zhao, X., Price, L. K., High, W., Suhl, J., & Cerel-Suhl, S. (2022). Sharpening Working Memory With Real-Time Electrophysiological Brain Signals: Which Neurofeedback Paradigms Work? *Front Aging Neurosci*, *14*, 780817. https://doi.org/10.3389/fnagi.2022.780817

Jirayucharoensak, S., Israsena, P., Pan-Ngum, S., Hemrungrojn, S., & Maes, M. (2019). A game-based neurofeedback training system to enhance cognitive performance in healthy elderly subjects and in patients with amnestic mild cognitive impairment. *Clin Interv Aging*, *14*, 347-360. https://doi.org/10.2147/CIA.S189047

Joosten-Weyn Banningh, L. W., Prins, J. B., Vernooij-Dassen, M. J., Wijnen, H. H., Olde Rikkert, M. G., & Kessels, R. P. (2011). Group therapy for patients with mild cognitive impairment and their significant others: results of a waiting-list controlled trial. *Gerontology*, *57*(5), 444-454. https://doi.org/10.1159/000315933

Kahlson, M. A., & Colodner, K. J. (2015). Glial Tau Pathology in Tauopathies: Functional Consequences. *J Exp Neurosci*, *9*(Suppl 2), 43-50. https://doi.org/10.4137/JEN.S25515

Kalova, E., Vlcek, K., Jarolimova, E., & Bures, J. (2005). Allothetic orientation and sequential ordering of places is impaired in early stages of Alzheimer's disease: corresponding results in real space tests and computer tests. *Behav Brain Res*, *159*(2), 175-186. https://doi.org/10.1016/j.bbr.2004.10.016

Katsumoto, A., Takeuchi, H., & Tanaka, F. (2019). Tau Pathology in Chronic Traumatic Encephalopathy and Alzheimer's Disease: Similarities and Differences. *Front Neurol*, *10*, 980. https://doi.org/10.3389/fneur.2019.00980

Kim, J. H., Park, S., & Lim, H. (2021). Developing a virtual reality for people with dementia in nursing homes based on their psychological needs: a feasibility study. *BMC Geriatr*, *21*(1), 167. https://doi.org/10.1186/s12877-021-02125-w

Kim, K., Han, J. W., So, Y., Seo, J., Kim, Y. J., Park, J. H., Lee, S. B., Lee, J. J., Jeong, H. G., Kim, T. H., & Kim, K. W. (2017). Cognitive Stimulation as a Therapeutic Modality for Dementia: A Meta-Analysis. *Psychiatry Investig*, *14*(5), 626-639. https://doi.org/10.4306/pi.2017.14.5.626

Kim, O., Pang, Y., & Kim, J. H. (2019). The effectiveness of virtual reality for people with mild cognitive impairment or dementia: a meta-analysis. *BMC Psychiatry*, *19*(1), 219. https://doi.org/10.1186/s12888-019-2180-x

Kovacs, G. G. (2017). Tauopathies. *Handb Clin Neurol*, *145*, 355-368. https://doi.org/10.1016/B978-0-12-802395-2.00025-0

Krell-Roesch, J., Feder, N. T., Roberts, R. O., Mielke, M. M., Christianson, T. J., Knopman, D. S., Petersen, R. C., & Geda, Y. E. (2018). Leisure-Time Physical Activity and the Risk of Incident Dementia: The Mayo Clinic Study of Aging. *J Alzheimers Dis*, *63*(1), 149-155. https://doi.org/10.3233/JAD-171141

Lambeck, S., [Sanne Lambeck], & Lambeck, J., [Johan Lambeck]. (2020). Aquatic Therapy: a valuable intervention in neurological and geriatric physiotherapy. A narrative review. EWAC Medical. https://www.ewacmedical.com/wp-content/uploads/2021/02/2020-Lambeck-Lambeck-Aquatic-Therapy-a-valuable-intervention-in-neurological-and-geriatric-p

Lautenschlager, N. T., Cox, K., & Kurz, A. F. (2010). Physical activity and mild cognitive impairment and Alzheimer's disease. *Curr Neurol Neurosci Rep*, *10*(5), 352-358. https://doi.org/10.1007/s11910-010-0121-7

Lavy, Y., Dwolatzky, T., Kaplan, Z., Guez, J., & Todder, D. (2019). Neurofeedback Improves Memory and Peak Alpha Frequency in Individuals with Mild Cognitive Impairment. *Appl Psychophysiol Biofeedback*, *44*(1), 41-49. https://doi.org/10.1007/s10484-018-9418-0

Lavy, Y., Dwolatzky, T., Kaplan, Z., Guez, J., & Todder, D. (2021). Mild Cognitive Impairment and Neurofeedback: A Randomized Controlled Trial. *Front Aging Neurosci*, *13*, 657646. https://doi.org/10.3389/fnagi.2021.657646

Leung, P., Orrell, M., & Orgeta, V. (2015). Social support group interventions in people with dementia and mild cognitive impairment: a systematic review of the literature. *Int J Geriatr Psychiatry*, *30*(1), 1-9. https://doi.org/10.1002/gps.4166

Leyns, C. E. G., & Holtzman, D. M. (2017). Glial contributions to neurodegeneration in tauopathies. *Mol Neurodegener*, *12*(1), 50. https://doi.org/10.1186/s13024-017-0192-x

Li, X., Yang, X., & Sun, Z. (2020). Alpha rhythm slowing in a modified thalamo-cortico-thalamic model related with Alzheimer's disease. *PLoS One*, *15*(3), e0229950. https://doi.org/10.1371/journal.pone.0229950

Linnemann, A., & Fellgiebel, A. (2017). [Psychotherapy with mild cognitive impairment and dementia]. *Nervenarzt*, *88*(11), 1240-1245. https://doi.org/10.1007/s00115-017-0408-x (Psychotherapie bei leichter kognitiver Beeintrachtigung und Demenz.)

Longhurst, J., Phan, J., Chen, E. H., Jackson, S. J., & Landers, M. R. (2020). Physical Therapy for Gait, Balance, and Cognition in Individuals with Cognitive Impairment: A Retrospective Analysis. Rehabilitation Research and Practice. https://doi.org/10.1155/2020/8861004

Luijmes, R. E., Pouwels, S., & Boonman, J. (2016). The effectiveness of neurofeedback on cognitive functioning in patients with Alzheimer's disease: Preliminary results. *Neurophysiol Clin*, *46*(3), 179-187. https://doi.org/10.1016/j.neucli.2016.05.069

Lyu, J., Zhang, J., Mu, H., Li, W., Champ, M., Xiong, Q., Gao, T., Xie, L., Jin, W., Yang, W., Cui, M., Gao, M., & Li, M. (2018). The Effects of Music Therapy on Cognition, Psychiatric Symptoms, and Activities of Daily Living in Patients with Alzheimer's Disease. *J Alzheimers Dis*, *64*(4), 1347-1358. https://doi.org/10.3233/JAD-180183

Magee, W. L., Ghetti, C. M., & Moyer, A. (2015). Feasibility of the music therapy assessment tool for awareness in disorders of consciousness (MATADOC) for use with pediatric populations. *Front Psychol*, *6*, 698. https://doi.org/10.3389/fpsyg.2015.00698

Mahendran, R., Gandhi, M., Moorakonda, R. B., Wong, J., Kanchi, M. M., Fam, J., Rawtaer, I., Kumar, A. P., Feng, L., & Kua, E. H. (2018). Art therapy is associated with sustained

Page 34 of 40

improvement in cognitive function in the elderly with mild neurocognitive disorder: findings from a pilot randomized controlled trial for art therapy and music reminiscence activity versus usual care. *Trials*, *19*(1), 615. https://doi.org/10.1186/s13063-018-2988-6

Marlats, F., Bao, G., Chevallier, S., Boubaya, M., Djabelkhir-Jemmi, L., Wu, Y. H., Lenoir, H., Rigaud, A. S., & Azabou, E. (2020). SMR/Theta Neurofeedback Training Improves Cognitive Performance and EEG Activity in Elderly With Mild Cognitive Impairment: A Pilot Study. *Front Aging Neurosci*, *12*, 147. https://doi.org/10.3389/fnagi.2020.00147

Marlats, F., Djabelkhir-Jemmi, L., Azabou, E., Boubaya, M., Pouwels, S., & Rigaud, A.-S. (2019). Comparison of effects between SMR/delta-ratio and beta1/theta-ratio neurofeedback training for older adults with Mild Cognitive Impairment: a protocol for a randomized controlled trial. *Trials*, *20*(1), 88. https://doi.org/10.1186/s13063-018-3170-x

Marques, C. L. (2019, May 27). Physical therapy in patients with Alzheimer's disease: a systematic review of randomized controlled clinical trials. Scielo.br. https://www.scielo.br/j/fp/a/hBBV8cDf9DS3tq7zJPTyxRR/?lang=en&format=pdf

Mendoza Laiz, N., Del Valle Diaz, S., Rioja Collado, N., Gomez-Pilar, J., & Hornero, R. (2018). Potential benefits of a cognitive training program in mild cognitive impairment (MCI). *Restor Neurol Neurosci*, *36*(2), 207-213. https://doi.org/10.3233/RNN-170754

Merino-Serrais, P., Benavides-Piccione, R., Blazquez-Llorca, L., Kastanauskaite, A., Rabano, A., Avila, J., & DeFelipe, J. (2013). The influence of phospho-tau on dendritic spines of cortical pyramidal neurons in patients with Alzheimer's disease. *Brain*, *136*(Pt 6), 1913-1928. https://doi.org/10.1093/brain/awt088

Page 35 of 40

Miall, R. C., Keating, J. G., Malkmus, M., & Thach, W. T. (1998). Simple spike activity predicts occurrence of complex spikes in cerebellar Purkinje cells. *Nat Neurosci*, *1*(1), 13-15. https://doi.org/10.1038/212

Miao, Y., Jurica, P., Struzik, Z. R., Hitomi, T., Kinoshita, A., Takahara, Y., Ogawa, K., Hasegawa, M., & Cichocki, A. (2021). Dynamic theta/beta ratio of clinical EEG in Alzheimer's disease. *J Neurosci Methods*, *359*, 109219. https://doi.org/10.1016/j.jneumeth.2021.109219

Miller, M. D., & Reynolds, C. F., 3rd. (2007). Expanding the usefulness of Interpersonal Psychotherapy (IPT) for depressed elders with co-morbid cognitive impairment. *Int J Geriatr Psychiatry*, *22*(2), 101-105. https://doi.org/10.1002/gps.1699

Moreno-Morales, C., Calero, R., Moreno-Morales, P., & Pintado, C. (2020). Music Therapy in the Treatment of Dementia: A Systematic Review and Meta-Analysis. *Front Med (Lausanne)*, *7*, 160. https://doi.org/10.3389/fmed.2020.00160

Moretti, D. V., Paternico, D., Binetti, G., Zanetti, O., & Frisoni, G. B. (2013). EEG upper/low alpha frequency power ratio relates to temporo-parietal brain atrophy and memory performances in mild cognitive impairment. *Front Aging Neurosci*, *5*, 63. https://doi.org/10.3389/fnagi.2013.00063

Nicodemus, N. E., Becerra, S., Kuhn, T. P., Packham, H. R., Duncan, J., Mahdavi, K., Iovine, J., Kesari, S., Pereles, S., Whitney, M., Mamoun, M., Franc, D., Bystritsky, A., & Jordan, S. (2019). Focused transcranial ultrasound for treatment of neurodegenerative dementia. *Alzheimers Dement (N Y)*, *5*, 374-381. https://doi.org/10.1016/j.trci.2019.06.007

Nicolae, I. E., & Ivanovici, M. (2021, 2-3 Sept. 2021). Complexity of EEG Brain Signals Triggered by Fractal Visual Stimuli. 2021 International Aegean Conference on Electrical Machines and Power Electronics (ACEMP) & 2021 International Conference on Optimization of Electrical and Electronic Equipment (OPTIM),

Nobukawa, S., Yamanishi, T., Nishimura, H., Wada, Y., Kikuchi, M., & Takahashi, T. (2019). Atypical temporal-scale-specific fractal changes in Alzheimer's disease EEG and their relevance to cognitive decline. *Cogn Neurodyn*, *13*(1), 1-11. https://doi.org/10.1007/s11571-018-9509-x

Okuda, M., Fujita, Y., Takada-Takatori, Y., Sugimoto, H., & Urakami, K. (2020). Aromatherapy improves cognitive dysfunction in senescence-accelerated mouse prone 8 by reducing the level of amyloid beta and tau phosphorylation. *PLoS One*, *15*(10), e0240378. https://doi.org/10.1371/journal.pone.0240378

Oliveira, J., Gamito, P., Souto, T., Conde, R., Ferreira, M., Corotnean, T., Fernandes, A., Silva, H., & Neto, T. (2021). Virtual Reality-Based Cognitive Stimulation on People with Mild to Moderate Dementia due to Alzheimer's Disease: A Pilot Randomized Controlled Trial. *Int J Environ Res Public Health*, *18*(10). https://doi.org/10.3390/ijerph18105290

Orgeta, V., Qazi, A., Spector, A. E., & Orrell, M. (2014). Psychological treatments for depression and anxiety in dementia and mild cognitive impairment. *Cochrane Database Syst Rev*(1), CD009125. https://doi.org/10.1002/14651858.CD009125.pub2

Ozbek, Y., Fide, E., & Yener, G. G. (2021). Resting-state EEG alpha/theta power ratio discriminates early-onset Alzheimer's disease from healthy controls. *Clin Neurophysiol*, *132*(9), 2019-2031. https://doi.org/10.1016/j.clinph.2021.05.012

Paddick, S. M., Mkenda, S., Mbowe, G., Kisoli, A., Gray, W. K., Dotchin, C. L., Ternent, L., Ogunniyi, A., Kissima, J., Olakehinde, O., Mushi, D., & Walker, R. W. (2017). Cognitive stimulation therapy as a sustainable intervention for dementia in sub-Saharan Africa: feasibility and clinical efficacy using a stepped-wedge design. *Int Psychogeriatr*, *29*(6), 979-989. https://doi.org/10.1017/S1041610217000163

Pais, R., Ruano, L., O, P. C., & Barros, H. (2020). Global Cognitive Impairment Prevalence and Incidence in Community Dwelling Older Adults-A Systematic Review. *Geriatrics (Basel)*, *5*(4). https://doi.org/10.3390/geriatrics5040084

Palmer, L. M., Clark, B. A., Grundemann, J., Roth, A., Stuart, G. J., & Hausser, M. (2010). Initiation of simple and complex spikes in cerebellar Purkinje cells. *J Physiol*, *588*(Pt 10), 1709-1717. https://doi.org/10.1113/jphysiol.2010.188300

Picken, C., Clarke, A. R., Barry, R. J., McCarthy, R., & Selikowitz, M. (2020). The Theta/Beta Ratio as an Index of Cognitive Processing in Adults With the Combined Type of Attention Deficit Hyperactivity Disorder. *Clin EEG Neurosci*, *51*(3), 167-173. https://doi.org/10.1177/1550059419895142

Popescu, T., Pernet, C., & Beisteiner, R. (2021). Transcranial ultrasound pulse stimulation reduces cortical atrophy in Alzheimer's patients: A follow-up study. *Alzheimers Dement (N Y)*, *7*(1), e12121. https://doi.org/10.1002/trc2.12121

Radinskaia, D., & Radinski, C. (2022). EEG coherence as a marker of Alzheimer's disease. *medRxiv*, 2022.2007.2024.22277966. https://doi.org/10.1101/2022.07.24.22277966

Raichle, M. E., & Snyder, A. Z. (2007). A default mode of brain function: a brief history of an evolving idea. *Neuroimage*, *37*(4), 1083-1090; discussion 1097-1089. https://doi.org/10.1016/j.neuroimage.2007.02.041

Reis, J., Portugal, A. M., Fernandes, L., Afonso, N., Pereira, M., Sousa, N., & Dias, N. S. (2016). An Alpha and Theta Intensive and Short Neurofeedback Protocol for Healthy Aging Working-Memory Training. *Front Aging Neurosci*, *8*, 157. https://doi.org/10.3389/fnagi.2016.00157

Rodriguez, G., Arnaldi, D., & Picco, A. (2011). Brain functional network in Alzheimer's disease: diagnostic markers for diagnosis and monitoring. *Int J Alzheimers Dis*, *2011*, 481903. https://doi.org/10.4061/2011/481903

Rolls, E. T. (2021). The connections of neocortical pyramidal cells can implement the learning of new categories, attractor memory, and top-down recall and attention. *Brain Struct Funct*, *226*(8), 2523-2536. https://doi.org/10.1007/s00429-021-02347-z

Rostamzadeh, A., Kahlert, A., Kalthegener, F., & Jessen, F. (2022). Psychotherapeutic interventions in individuals at risk for Alzheimer's dementia: a systematic review. *Alzheimers Res Ther*, *14*(1), 18. https://doi.org/10.1186/s13195-021-00956-8

Salehpour, F., Khademi, M., & Hamblin, M. R. (2021). Photobiomodulation Therapy for Dementia: A Systematic Review of Pre-Clinical and Clinical Studies. *J Alzheimers Dis*, *83*(4), 1431-1452. https://doi.org/10.3233/JAD-210029

Saltmarche, A. E., Naeser, M. A., Ho, K. F., Hamblin, M. R., & Lim, L. (2017). Significant Improvement in Cognition in Mild to Moderately Severe Dementia Cases Treated with Transcranial Plus Intranasal Photobiomodulation: Case Series Report. *Photomed Laser Surg*, *35*(8), 432-441. https://doi.org/10.1089/pho.2016.4227

Sauzeon, H., N'Kaoua, B., Pala, P. A., Taillade, M., Auriacombe, S., & Guitton, P. (2016). Everyday-like memory for objects in ageing and Alzheimer's disease assessed in a visually complex environment: The role of executive functioning and episodic memory. *J Neuropsychol*, *10*(1), 33-58. https://doi.org/10.1111/jnp.12055

Schmidt, M. T., Kanda, P. A., Basile, L. F., da Silva Lopes, H. F., Baratho, R., Demario, J. L., Jorge, M. S., Nardi, A. E., Machado, S., Ianof, J. N., Nitrini, R., & Anghinah, R. (2013). Index of alpha/theta ratio of the electroencephalogram: a new marker for Alzheimer's disease. *Front Aging Neurosci*, *5*, 60. https://doi.org/10.3389/fnagi.2013.00060

Se Thoe, E., Fauzi, A., Tang, Y. Q., Chamyuang, S., & Chia, A. Y. Y. (2021). A review on advances of treatment modalities for Alzheimer's disease. *Life Sci*, *276*, 119129. https://doi.org/10.1016/j.lfs.2021.119129

Sherratt, K., Thornton, A., & Hatton, C. (2004). Music interventions for people with dementia: a review of the literature. *Aging Ment Health*, *8*(1), 3-12. https://doi.org/10.1080/13607860310001613275

Smits, F. M., Porcaro, C., Cottone, C., Cancelli, A., Rossini, P. M., & Tecchio, F. (2016). Electroencephalographic Fractal Dimension in Healthy Ageing and Alzheimer's Disease. *PLoS One*, *11*(2), e0149587. https://doi.org/10.1371/journal.pone.0149587

Styliadis, C., Kartsidis, P., Paraskevopoulos, E., Ioannides, A. A., & Bamidis, P. D. (2015). Neuroplastic effects of combined computerized physical and cognitive training in elderly individuals at risk for dementia: an eLORETA controlled study on resting states. *Neural Plast*, *2015*, 172192. https://doi.org/10.1155/2015/172192

Yan, Y. J., Lin, R., Zhou, Y., Luo, Y. T., Cai, Z. Z., Zhu, K. Y., & Li, H. (2021). Effects of expressive arts therapy in older adults with mild cognitive impairment: A pilot study. *Geriatr Nurs*, *42*(1), 129-136. https://doi.org/10.1016/j.gerinurse.2020.11.011

Yang, L., Wu, C., Parker, E., Li, Y., Dong, Y., Tucker, L., Brann, D. W., Lin, H. W., & Zhang, Q. (2022). Non-invasive photobiomodulation treatment in an Alzheimer Disease-like transgenic rat model. *Theranostics*, *12*(5), 2205-2231. https://doi.org/10.7150/thno.70756

Zempolich, G. W., Brown, S. T., Holla, M., & Raman, I. M. (2021). Simple and complex spike responses of mouse cerebellar Purkinje neurons to regular trains and omissions of somatosensory stimuli. *J Neurophysiol*, *126*(3), 763-776. https://doi.org/10.1152/jn.00170.2021

Zucchella, C., Sinforiani, E., Tamburin, S., Federico, A., Mantovani, E., Bernini, S., Casale, R., & Bartolo, M. (2018). The Multidisciplinary Approach to Alzheimer's Disease and Dementia. A Narrative Review of Non-Pharmacological Treatment. *Front Neurol*, *9*, 1058. https://doi.org/10.3389/fneur.2018.01058

Zueva, M. V. (2015). Fractality of sensations and the brain health: the theory linking neurodegenerative disorder with distortion of spatial and temporal scale-invariance and fractal complexity of the visible world. *Front Aging Neurosci*, *7*, 135. https://doi.org/10.3389/fnagi.2015.00135